\documentstyle[11pt]{article}
\begin{document}
\parskip=2mm
\topmargin -10mm
\pagestyle{plain}

\begin{tabbing}\hspace{12cm}\={\small preprint UTF-397}\\
\> gr-qc/9704077\\
\> April 1997
\end{tabbing}
\vspace{1cm}

\begin{center}
{\Large\bf Massive scalar field
near a cosmic string}
\vspace{5mm}

{\Large Devis Iellici\footnote{Electronic address:
iellici@science.unitn.it}} \\
\vspace{3mm}

{Dipartimento di Fisica, Universit\`a di Trento\\
and Istituto Nazionale di Fisica Nucleare,\\
Gruppo Collegato di Trento, Italia} 
\end{center}
\vspace{1cm}
\begin{center}
\parbox{13cm}{\small{\bf Abstract.} 
The $\zeta$ function of a massive scalar field near a cosmic string is
computed and then employed to find the vacuum fluctuation of the
field. The vacuum expectation value of the energy-momentum tensor is
also computed using a point-splitting approach. The obtained results
could be useful also for the case of self-interacting scalar fields
and for the finite-temperature  Rindler space theory.}
\end{center}

\hspace{10mm}{\small PACS number(s): 04.62.+v, 04.70.Dy }

\vspace{1cm}

\begin{center}
{\large\bf I. Introduction}
\end{center}

In the last few years many authors have considered quantum fields on
spacetimes containing conical singularities. These kind of manifolds
are relevant for studying fields in presence of real conical
singularities, namely existing in the Lorentzian section of the
manifold, as in the case of idealized cosmic strings and of
2+1-dimensional gravity. In other cases,  conical singularities appear
in the finite temperature theory, when adopting the Euclidean path
integral with periodic imaginary time formalism. It is the case of
spacetimes containing horizons, such as the Rindler and the
Schwarzschild spaces, which are much studied in the off-shell 
approach to the one-loop quantum corrections to the Bekenstein-Hawking
entropy. It is worth noticing that, in these latter cases, the conical
singularity could be avoided adopting a canonical approach.

As a consequence of this interest in conical spaces, many techniques
have been developed to compute relevant quantities in presence of
conical singularities (see, among the others,
\cite{DO77,cheeger83,DO87a,DesJack88,AlOt90,GiRuVa90,KS91,CKV94,fursaev94,FUSOL95,ZCV}).
With few exceptions 
\cite{CKV94,fursaev94,linet87,ShiraHire92,GuimLin93,GuimLin94,moreira95},
most authors considered massless fields only. Indeed, the
introduction of the mass complicates considerably the problem, so that
it becomes very difficult to obtain manageable forms for the physical
quantities. At this regard, it has to be noted that using the
heat-kernel approach \cite{CKV94,fursaev94} it seems that all the
problems vanish: in fact, for a scalar field of mass $m$ the
integrated heat-kernel  is simply related to the massless one as
\begin{eqnarray}
 K_t^{m}=e^{-tm^2}K_t^{m=0},
\label{heat}
\end{eqnarray}
and, at least for a simple cone $R^{D-2}\times C_\alpha$,
the integrated massless heat-kernel is well known
\cite{cheeger83,CKV94,fursaev94}:
\begin{eqnarray*}
K_t^{m=0}=\frac{\Sigma_{D-2}}{(4\pi t)^2}\left[\frac{V(C_\alpha)}
{4\pi t}+\frac{1}{12}\left(\frac{2\pi}{\alpha}-\frac{\alpha}{2\pi}
\right)\right],
\end{eqnarray*}
where $2\pi-\alpha$ is the deficit angle of the cone, $V(C_\alpha)$ is
the volume of the cone and $\Sigma_{D-2}$ is the volume of the
transverse dimensions. On the contrary of the massless case, the
Mellin transform of Eq. (\ref{heat}) exists in the massive case, and
so it is easy to compute the massive global $\zeta$ function and from
it quantities like the effective action, which have a simple closed
form. However, it is well known \cite{ZCV} that this procedure does
not yield the correct dependence on the deficit angle of the cone in
the massless limit and, moreover, these global quantities are finite,
apart for the volume divergences, while the local quantities show a
non-integrable singularity near the tip of the cone.\footnote{In the
cosmic string case, this problem could be overcome renormalizing the
action of the string as in \cite{fursaev94}.} Therefore, it is at
least dubious that this simple result for the massive case is correct.
Besides, in order to compute important quantities such as the vacuum
fluctuation of the field  and the stress tensor, local rather than
global quantities are needed.

The aim of this work is to give a more manageable tool to compute
local quantities for a massive scalar field around a cosmic string,
and we will accomplish this by computing the local $\zeta$ function,
which will be given as an expansion in powers of $mr$, where $r$ is
the distance form the string core. Therefore, the obtained expression
is useful near the string, where $mr\ll 1$.\footnote{Note that
$r=\hbar/mc$ is the Compton wavelength of the particle. When the
result are translated for the Rindler case, the approximation is valid
close to the event horizon.} However,
since the local quantities diverge at the string core, this is just the
interesting region if, for instance, one wants to consider the
back-reaction of the energy-momentum tensor on the background 
metric \cite{Hiscock87,MEX97}. The expression we obtain is simple
enough to allow us to compute the renormalized vacuum expectation
value $\langle \phi^2(x)\rangle$ and the the energy-momentum tensor,
potentially up to arbitrary order in $(mr)^2$.
Notice that the most common tool used to compute local quantities is
the Green's function, which in the massive case is given in a
complicated integral  form \cite{linet87,ShiraHire92,GuimLin94,CKV94} 
or as a sum of generalized hypergeometric functions 
\cite{moreira95}.

The capability of treating the massive case is  useful also when
dealing with self-interacting theories, such as $\lambda \phi^4$. In
fact, consider an Euclidean path-integral approach to the theory of a
self-interacting scalar field: the generating functional of the theory
is the given by  
\begin{eqnarray*}
Z&=&\int{\cal{D}}\phi\exp(-S[\phi]),\\
S[\phi]&=&\int d^Dx\sqrt{g}\left[\frac{1}{2}\partial_\mu\phi 
\partial^\mu\phi+V(\phi)\right],
\end{eqnarray*}
where $V(\phi)$ is a potential. In the usual $\phi^4$-theory, it has
the form $V(\phi)=\frac{1}{2}m^2\phi^2+\frac{1}{4!}\lambda\phi^4$,
$\lambda$ being the coupling constant. In the one-loop perturbative
approach, the above action is expanded near the minimum $\hat\phi$ 
of the action up to second order terms
$$
S[\hat\phi+\varphi]=S[\hat\phi]+\frac{1}{2}\int d^Dx\sqrt{g}\,
\varphi[-\Delta+M^2]\varphi,
$$
where $\Delta$ is the Laplace-Beltrami operator and
$M^2=V''(\hat\phi)$. In the $\phi^4$-case, we have that
$M^2=m^2+\frac{1}{2}\lambda{\hat\phi}^2$.
Then, the standard step is to assume $\hat\phi$ is a constant 
configuration, so that standard passages lead to effective
action 
$$
W=-\ln Z=S[\hat\phi]+\frac{\hbar}{2}\ln\det[-\Delta+M^2]+
{\cal O}(\hbar^2),
$$
where, in general, $S[\hat\phi]$ is proportional to the volume of the
spacetime. The determinant in the second term is ultraviolet
divergent and it has to be defined by means of some regularization
procedure. Convenient ways to regularize the above determinant are,
for example, the heat-kernel plus Schwinger's proper-time
regularization or the $\zeta$-function regularization.

{}From the above outline, it is clear that the one-loop self-interacting
case is closely related to the massive case, and, in particular, the
regularization of the determinant is the same, provided that $m^2$
is replaced by $M^2$. Of course, the renormalization of the effective
action to remove the ultraviolet divergences is different. In this
paper we shall not discuss the self-interacting case and its
renormalization explicitly, but we observe that in the cosmic string
case, in which $\langle \phi^2(x)\rangle\propto r^{-2}$, it could be
more reliable a `Hartree-like' approximation as that employed in
\cite{ShiraHire92} and in which in the action $\lambda\phi^4$ is
replaced by $\lambda\langle \phi^2\rangle\phi^2$.

As we said above, we are mainly concerned in the comic-string
background. We remind that the space-time around an infinitely long,
static and straight cosmic string \cite{vilenk85} has the topology
${\cal M}^4=R^2\times C_\alpha$, where $C_\alpha$ is the simple
two-dimensional cone with deficit angle $2\pi-\alpha$, and the metric
is
\begin{eqnarray}
ds^2=-dt^2+dz^2+dr^2+r^2 d\theta^2,\hspace{1cm} r\in(0,+\infty),\,
\,\theta\in[0,\alpha],\, \,t,z\in(-\infty,+\infty).
\label{metric}
\end{eqnarray}
The polar angle deficit $2\pi-\alpha$ is related to the mass per unit
length of the string $\mu$ by $2\pi-\alpha=8\pi G\mu$. Throughout this
paper we shall assume positive deficit angle, so that
$\nu\equiv2\pi/\alpha>1$. For GUT strings $\nu-1\sim 10^{-6}$
\cite{vilenk85}. In the rest of the paper we  adopt an Euclidean
approach and so we perform a Wick rotation $\tau=it$: the form of the
metric is the same as above with the replacement
$-dt^2\rightarrow+d\tau^2$. Moreover, for technical reasons we will
also consider the obvious generalization to the $D$-dimensional case,
with topology $R^{D-2}\times C_\alpha$.

We have assumed zero thickness for the string: this is clearly an
idealization, since an actual cosmic string should have a small  but
finite radius, of the order of the Compton wavelength of the Higgs
boson involved in the phase transition which gives rise to the cosmic
string \cite{vilenk85}. The internal structure of the string has
non-negligible effects even at large distances, as it has been shown
by Allen, Kay, and Ottewill \cite{AlKaOt96}. However, in the same work
it has been shown that for a minimally coupled scalar this dependence
on the internal structure is absent, and one can safely use the
idealized string. In order to avoid the complications discussed in
\cite{AlKaOt96}, which are related to the non-uniqueness of the
self-adjoint extension of the Laplace-Beltrami operator in the
idealized conical space-time, we shall consider the minimally coupled
case only. Actually, in Sec. IV we will compute the stress tensor 
and, for sake of completeness, we will consider arbitrary coupling.
Therefore,  we have to remember that for $\xi\neq 0$ others effects
could be present in a realistic cosmic string.

Although the results we will obtain are easily translated to the 
case of fields at finite temperature in the Rindler space
\footnote{The Euclidean metric of the cosmic string and 
of the finite-temperature Rindler space are in fact the same.
In this latter case, $\theta$ corresponds to the periodic
Rindler time and the period $\alpha$ is the inverse temperature.
The absence of the conical singularity fixes the Unruh-Hawking
temperature $T=1/2\pi$ in suitable units \cite{birrel}.}
or near the horizon of a Schwarzschild black hole, we prefer
to consider the cosmic string background only. This because 
the Euclidean Rindler space or Schwarzschild black hole
show a conical singularity only when the temperature is 
different from the Unruh-Hawking one,  namely when one
considers the `off-shell' theories, but it has been shown that the
off-shell quantum states of a field are affected by several
pathologies on the event horizon \cite{HaNaSt84} and that these
theories do not have a consistent thermodynamics \cite{moiel}.

The rest of this work is organized as follows. In Sec. II we compute
the $\zeta$ function of a massive scalar field in the cosmic string
background. Then we use this result to compute, in the region $mr\ll
1$, the vacuum fluctuation of the field in Sec. III and the expectation
value of the stress tensor in Sec. IV. Section V contains the
conclusions. In the rest of the the paper we use units
for which $\hbar=c=G=1$.

\begin{center}
{\large\bf II. Computation on the $\zeta$ function}
\end{center}

We consider a quasi-free real scalar field in the background given by
the (Wick-rotated) metric (\ref{metric}). The action of the theory is
the given by
\begin{eqnarray*}
S[\phi]=\int d^Dx\sqrt{g}\frac{1}{2}\phi A\phi,
\end{eqnarray*}
where $A$ is known as the small fluctuation operator and in our
case reads
\begin{eqnarray}
A=-\Delta_D+m^2+\xi R.
\end{eqnarray}
Here $\xi$ is a parameter which fixes the coupling of the field to the
gravity by means of the scalar curvature $R$. We shall consider the
minimally coupled case, $\xi=0$. Then, by means standard passages 
one sees that the generating functional of the theory can be expressed
in terms of the functional determinant of the small fluctuations
operator, which can be conveniently defined by means of the $\zeta$
function regularization \cite{hawking77}:
$$
\ln Z_\alpha=-\frac{1}{2}\ln\det\mu^{-2}A=\frac{1}{2}
\zeta'(0|A\mu^{-2})
$$
where $\zeta(s|A)$ is the global $\zeta$ function related to the
operator $A$, the prime indicates the derivative with respect to $s$
and $\mu$ is an arbitrary parameter with the dimensions of a mass
needed for dimensional reasons and not to be confused with the mass
per unit length of the string. The global $\zeta$ function can be
formally written as the integral over the manifold of a local $\zeta$
function $\zeta(s|A)(x)$:
\begin{eqnarray*}
\zeta(s|A)&=&\int_{{\cal M}^D_\alpha}\zeta(s|A)(x)\sqrt{g} d^Dx.
\end{eqnarray*}
Actually, when the manifold is non-compact only the local $\zeta$
function has a precise mathematical meaning, since the integration
requires the introduction of cutoffs or smearing functions to avoid
divergences.

In the massless case, the $\zeta$ function on the cone has been
explicitly computed in \cite{ZCV}. In order to compute it in the
massive case, the starting point is the well known relation between
the local $\zeta$ function and the heat kernel given by the Mellin
transform:
\begin{eqnarray*}
\zeta^{m}(s;x)=
\frac{1}{\Gamma(s)}\int_0^\infty dt\,t^{s-1}e^{-tm^2}K_t^{m=0}(x).
\end{eqnarray*}
The heat kernel of the massive field is related to the massless
one by means of the following obvious decomposition:
\begin{eqnarray*}
K_t^{m}(x)=e^{-tm^2}K_t^{m=0}(x).
\end{eqnarray*}
Now we take into account the following property of the Mellin
transform of the product of two functions \cite{report}
\begin{eqnarray*}
\int_0^\infty t^{s-1}f(t)g(t)dt=
\frac{1}{2\pi i}\int_{\sigma-i\infty}^{\sigma-i\infty}
F(z)G(s-z)dz,
\end{eqnarray*}
where $F$ and $G$ are the Mellin transforms of $f$ and $g$
respectively and $\sigma$ is a real number in the common strip of
convergence of the two  Mellin transforms. In this way we can write
the massive $\zeta$ function of  in terms of the massless one:
\begin{eqnarray}
\zeta^{m}(s;x)=\frac{1}{2\pi i \Gamma(s)}
\int_{\sigma-i\infty}^{\sigma-i\infty}
\Gamma(z)\zeta^{m=0}(z;x)m^{2z-2s}\Gamma(s-z) dz,
\label{general}
\end{eqnarray}
where we have used the Mellin transform
\begin{eqnarray*}
\int_0^\infty t^{s-1}e^{-tm^2}dt=m^{-2s}\Gamma(s),
\end{eqnarray*}
which converges for $\mbox{Re} s>0$.

Since we are interested in the $\zeta$ function of a massive scalar
field in the (Euclidean) space $ R^{D-2}\times C_\alpha$, we consider
the corresponding massless $\zeta$ function. It has been shown in
\cite{cheeger83} and \cite{ZCV} that it is the sum of two parts which
converge in separate strips of the complex $s$-plane and which are
summed only after their analytic continuation:
\begin{eqnarray*}
\zeta^{m=0}(z;x)=\zeta_<(s;x)+\zeta_>(s;x),
\end{eqnarray*}
where 
\begin{eqnarray*}
\zeta_<(s;x)&=&\frac{r^{2s-D}}{(4\pi)^{\alpha\frac{D-2}{2}}
\Gamma(s)} \frac{\Gamma(s-\frac{D-1}{2})
\Gamma(\frac{D}{2}-s)}{2\sqrt{\pi}\Gamma(s-\frac{D-2}{2})}, 
\hspace{1cm}\frac{D-1}{2}<\mbox{Re} s<\frac{D}{2},\\
\zeta_>(s;x)&=&\frac{r^{2s-D}}{\alpha(4\pi)^{\frac{D-2}{2}}
\Gamma(s)} \frac{\Gamma(s-\frac{D-1}{2})}{\sqrt{\pi}}
G_\alpha(s-{\scriptstyle\frac{D-2}{2}}),
\hspace{1cm}
\frac{D}{2}<\mbox{Re} s<\frac{D}{2}+\nu,
\end{eqnarray*}
and the function $G_\alpha(s)$ is defined as
\begin{eqnarray*}
G_\alpha(s)=\sum_{n=1}^{\infty}
\frac{\Gamma(\nu_n-s+1)}{\Gamma(\nu_n+s)},
\hspace{1cm}\nu_n=\frac{2\pi}{\alpha}|n|,\hspace{0.5cm}
\nu\equiv \nu_1
\end{eqnarray*}
and can be analytically continued in the whole complex plane showing
simple poles in $s=1$, with residue $\alpha/4\pi$, and $s=\nu_n+k+1$,
$k=0,1,2,\dots,$ (if $\alpha\neq2\pi$) with obvious residue. Since
$\zeta_<$ and $\zeta_>$ do not have a common strip of convergence, 
we must split Eq. (\ref{general}) in two parts: setting $\mbox{Re}
s>D/2$ we have
\begin{eqnarray*}
\zeta_>^{m}(s;x)&=&\frac{1}{2\pi i\Gamma(s)}
\frac{r^{-D}m^{-2s}}{(4\pi)^{\frac{D-2}{2}}\sqrt{\pi}\alpha}
\int_{\scriptsize \frac{D}{2}<{\mbox{\scriptsize Re}\,} z<
{\mbox{\scriptsize Re}\,} s}
\hspace{-8mm}dz\,\Gamma(z-{\scriptstyle\frac{D-1}{2}})
G_\alpha(z-{\scriptstyle\frac{D-2}{2}}) \Gamma(s-z)(mr)^{2z} 
\nonumber\\
&=&\frac{r^{2s-D}}{(4\pi)^{\frac{D-2}{2}}\sqrt{\pi}\alpha\Gamma(s)}
\left\{\sum_{n=0}^\infty\frac{(-1)^n}{n!}(mr)^{2n}
\Gamma(s+n-{\scriptstyle\frac{D-1}{2}})
G_\alpha(s+n-{\scriptstyle\frac{D-2}{2}})\right.
\nonumber\\
&&+\left.(mr)^{D-2s}
\sum_{n=1}^\infty\sum_{k=0}^\infty\frac{(-1)^k}{k!}
\frac{\Gamma(\nu_n+k+1/2)}{\Gamma(2\nu_n+k+1)}
\Gamma(s-\nu_n-k-{\scriptstyle\frac{D}{2}})
(mr)^{2\nu_n+2k}\right\},
\end{eqnarray*}
where we have performed the integral shifting the integration path to
the right and picking up the residues of the poles of the integrand.
The same procedure can be applied to $\zeta_<^{m}(s;x)$, but in this
case we have to consider the poles in $z=s+n$ and $z=2+n$:
\begin{eqnarray*}
\zeta_<^{m}(s;x)&=&\frac{1}{2\pi
i\Gamma(s)}\frac{r^{-D}m^{-2s}}{(4\pi)^{\frac{D-2}{2}}\sqrt{\pi}\alpha} 
\int_{\scriptsize\frac{D-1}{2}<{\mbox{\scriptsize Re}\,} z<
{\mbox{\scriptsize Re}\,} \frac{D}{2}}
\hspace{-5mm}dz\,\frac{\Gamma(z-\frac{D-1}{2})
\Gamma(\frac{D}{2}-z)}
{2\Gamma(z+1-D/2)}\Gamma(s-z)(mr)^{2z}\nonumber\\
&=&\frac{r^{2s-D}}{(4\pi)^{\frac{D-2}{2}}\sqrt{\pi}\alpha\Gamma(s)}
\left\{\sum_{n=0}^\infty\frac{(-1)^n}{n!}(mr)^{2n}
\frac{\Gamma(s+n-\frac{D-1}{2})\Gamma(\frac{D}{2}-s-n)}
{2\Gamma(s+n+1-D/2)}\right.
\nonumber\\ 
&&\left.+(mr)^{D-2s}\sum_{n=0}^\infty\frac{(-1)^n}{n!}(mr)^{2n}
\frac{\Gamma(n+1/2)\Gamma(s-n-D/2)}{2\Gamma(n+1)}\right\}.
\end{eqnarray*}
Now we analytically continue each term and sum to get the final
expression of the local $\zeta$ function of a massive scalar field:
\begin{eqnarray}
\zeta^{m}(s;x)&=&\frac{r^{2s-D}\mu^{2s}}{(4\pi)^{\frac{D-2}{2}}
\alpha\Gamma(s)}\left\{\sum_{n=0}^\infty\frac{(-1)^n}{n!}(mr)^{2n}
I_\alpha(s+n-{\scriptstyle\frac{D-2}{2}})
\right.\nonumber\\
&&\hspace{-11mm}+\frac{(mr)^{D-2s}}{\sqrt{\pi}}
\sum_{n=0}^\infty\frac{(-1)^n}{n!}(mr)^{2n}
\frac{\Gamma(n+1/2)\Gamma(s-n-D/2)}{2\Gamma(n+1)}\nonumber\\
&&\hspace{-11mm}\left.+\frac{(mr)^{D-2s}}{\sqrt{\pi}}
\sum_{n=1}^\infty\sum_{k=0}^\infty\frac{(-1)^k}{k!}
\frac{\Gamma(\nu_n+k+1/2)}{\Gamma(2\nu_n+k+1)}
\Gamma(s-\nu_n-k-D/2)
(mr)^{2\nu_n+2k}\right\},\nonumber\\
&&
\label{zetafun}
\end{eqnarray}
where the function $I_\alpha(s)$ is defined as \cite{ZCV}
\begin{eqnarray}
I_\alpha(s)&=&\frac{\Gamma(s-1/2)}{\sqrt{\pi}}
\left[G_\alpha(s)+\frac{\Gamma(1-s)}{2\Gamma(s)}\right],
\nonumber\\
I_\alpha(0)&=&\frac{1}{6\nu}\left(\nu^2-1\right),\nonumber\\
I_\alpha(-1)&=&\frac{1}{90\nu}\left(\nu^2-1\right)
\left(\nu^2+11\right), 
\label{ibeta}
\end{eqnarray}
The function $I_\alpha(s)$ is analytic in the whole complex plane but
in $s=1$, where it has a simple pole. Near the pole we have
$$
I_\alpha(s)=\frac{1}{2(s-1)}\left(\nu^{-1}-1\right)+
\frac{1}{2}\left(\nu^{-1}-1\right)\left(\gamma-2\ln 2\right)
-\frac{\ln\nu}{\nu}+{\cal {O}}\left((s-1)^2\right),
$$
where $\gamma$ is the Euler's constant. In expression (\ref{zetafun})
we have also reintroduced the arbitrary mass $\mu$ using the formal
relation $\zeta(s|A\mu^{-2})=\mu^{2s} \zeta(s|A)$, and which has been
omitted in the derivation for simplicity. It must be noted that the
above $\zeta$ function can be obtained in a more direct way, albeit
much longer, performing the integrations in the Mellin transform of
the spectral representation of the massive heat kernel, and then
rearranging the sums and the Euler gamma functions in the generalized
hypergeometric functions obtained by the integrations. We prefer the
used method since it can be more easily applied to the computation of
others quantities, as we will see in the next sections.

Although expression (\ref{zetafun}) looks awful, it is very simple
when $mr\ll 1$: considering the physically interesting case
$D=4$ and terms up to\footnote{Note that we are interested in the
values of the $\zeta$ function at $s=0,1$, and we consider positive
deficit angles only, so that $\nu>1$.} $(mr)^4$
\begin{eqnarray*}
\zeta^{m}(s;x)&=&\frac{r^{2s-4}\mu^{2s}}
{4\pi\alpha\Gamma(s)}\left[I_\alpha(s-1)-
(mr)^2 I_\alpha(s)+\frac{1}{2}(mr)^4 I_\alpha(s+1)+
{\cal O}\left((mr)^6\right)\right]\nonumber\\
&&+\frac{m^4(m/\mu)^{-2s}}{8\pi\alpha\Gamma(s)}
\left[\Gamma(s-2)-\frac{(mr)^2}{2}\Gamma(s-3)\right.\\
&&\left.+2(mr)^{2\nu}
\frac{\Gamma(\nu+1/2)\Gamma(s-\nu-2)}
{\sqrt{\pi}\Gamma(2\nu+1)}+{\cal O}\left((mr)^4\right)\right].
\end{eqnarray*}
In the first term of the first row we recognize the massless $\zeta$
function on $R^2\times C_\alpha$ \cite{ZCV}, while the first term in
the second row becomes the $\zeta$ function of a massive scalar field
in the Minkowski space-time when $\alpha=2\pi$ \cite{AAA95}. The
others terms are clearly corrections due to the presence of the
conical singularity. When the singularity is absent, namely when
$\alpha=2\pi$, the first row vanishes, since $I_{2\pi}(s)=0$, while in
the second and third rows only the first term survives, since the
others cancel two by two.

At the same order in $mr$, the effective lagrangian density is 
given by 
\begin{eqnarray*}
{\cal L}(x)&=&\frac{1}{2}\frac{d}{ds}\zeta^m(s;x)|_{s=0}\nonumber\\
&=&\frac{1}{8\pi\alpha r^4}\left\{I_\alpha(-1)
-(mr)^2 I_\alpha(0)\right.\nonumber\\
&&-\frac{(mr)^4}{2}
\left[\frac{\nu-1}{\nu}\left(\gamma+\ln\frac{{r\mu}}{2}\right)+
\frac{\ln\nu}{\nu}+\ln\frac{m}{\mu}-\frac{3}{4}\right]\\
&&\left.-\frac{(mr)^6}{24}\left[2G_\alpha(2)+2\gamma+
2\ln\frac{mr}{2}-\frac{11}{6}\right]+2(mr)^{2\nu+4}
\frac{\Gamma(\nu+\frac{1}{2})\Gamma(-\nu-2)}
{\sqrt{\pi}\Gamma(2\nu+1)}\right\}.
\end{eqnarray*}
where  $G_\alpha(n)$, $n\geq 2$, are readily computed being in the
region of convergence of the series defining $G_\alpha$. For example
\begin{eqnarray*}
G_\alpha(2)&=&-\frac{1}{2\nu}\left[2\gamma+
\psi\left(\nu^{-1}\right)+
\psi\left(-\nu^{-1}\right)\right],
\nonumber\\
G_\alpha(3)&=&-\frac{1}{6\nu}\left[3\gamma-
\psi\left(\nu^{-1}\right)-
3\psi\left(-\nu^{-1}\right)
+\psi\left(-2\nu^{-1}\right)\right],
\end{eqnarray*}
where $\psi(x)=\Gamma'(x)/\Gamma(x)$. In the limit 
$\alpha\rightarrow 2\pi$ the above effective lagrangian reduces to the
usual Coleman-Weinberg potential. It is important to note that the
computation of higher orders in the above expansion in powers of
$(mr)^2$ does not involve particular complications.

\begin{center}
{\large\bf III. Vacuum fluctuations}
\end{center}

Now we use the above $\zeta$ function to compute the (renormalized) value
of the vacuum fluctuation of the field \cite{AAA95}:
$$
\langle\phi^2(x)\rangle=\lim_{s\rightarrow 1}\zeta(s;x).
$$
Since even in the Minkowski space the vacuum expectation value
diverges, we renormalize the vacuum expectation value on the cone
subtracting the Minkowski value: in this way we have a remarkable
cancellation of the poles, yielding a finite result: 
\begin{eqnarray}
\langle\phi^2(x)\rangle_\alpha-\langle\phi^2(x)\rangle_{2\pi}&=&
\frac{1}{4\pi\alpha r^2} \left\{I_\alpha(0)+\frac{(mr)^2}{\nu}
\left[(\nu-1)\left(\ln\frac{mr}{2}+\gamma-\frac{1}{2}\right)+
\ln\nu\right]\right.\nonumber\\
&&+\frac{(mr)^4}{8}\left[2G_\alpha(2)+2\gamma-1+
2\ln\frac{mr}{2}\right]\nonumber\\
&&\left.+(mr)^{2\nu+2}
\frac{\Gamma(\nu+1/2)\Gamma(-\nu-1)}{\sqrt{\pi}\Gamma(2\nu+1)}+
{\cal O}\left((mr)^6\right)\right\}.
\label{vacuum}
\end{eqnarray}
The computation of higher corrections is straightforward. One can
verify that Eq. (\ref{vacuum}) correctly vanish in the limit
$\alpha\rightarrow 2\pi$, and that up to order $(mr)^2$ it is
identical to the result obtained by Moreira Jnr \cite{moreira95}. In
particular, we notice the additional logarithmic divergences of the
vacuum fluctuations at the conical singularity due to the massive
corrections.

\begin{center}
{\large\bf IV. Energy-momentum tensor}
\end{center}

Another important vacuum average is that of the energy-momentum
tensor. While in the massless case it is quite easy to be computed,
since its form is fixed from symmetry arguments, in the massive case
the computation is much more difficult and, to our knowledge, the
massive corrections near the string have never  been shown explicitly.
Only in  \cite{ShiraHire92} and \cite{GuimLin94} the explicit form of
the energy-momentum tensor has been given for $mr\gg 1$, where they
found the expected exponential damping factor $\exp(-2mr)$.

We use a point-splitting approach, in which the vacuum expectation
value of the energy momentum tensor is given by the coincidence 
limit of a non-local differential operator applied to the propagator 
of the field \cite{birrel}:
\begin{eqnarray*}
\langle T_{\mu\nu}(x)\rangle=i\lim_{x'\rightarrow x}D_{\mu\nu}(x,x')
G_{\cal F}(x,x'),
\end{eqnarray*}
where 
\begin{eqnarray}
D_{\mu\nu}(x,x')\equiv(1-2\xi)\nabla_\mu\nabla_{\nu'}+
-2\xi \nabla_\mu\nabla_\nu
+(2\xi-1/2)g_{\mu\nu}\left[\nabla_\alpha \nabla^{\alpha'}
-m^2\right],
\label{dimunu}
\end{eqnarray}
and the prime indicates that the derivative has to be taken with
respect to $x'$.   Since the propagator can be obtained from the
off-diagonal $\zeta$ function \cite{birrel} as
\begin{eqnarray*}
G_{\cal F}(x,x')=i\lim_{s\rightarrow 1}\zeta(s;x,x'),
\end{eqnarray*}
we can compute the energy momentum tensor from the 
$\zeta$ function as (see also \cite{hawking77,CVZ90})
\begin{eqnarray*}
\langle T_{\mu\nu}(x)\rangle=-\lim_{s\rightarrow 1}
\lim_{x'\rightarrow x}
D_{\mu\nu}(x,x')\zeta(s;x,x').
\end{eqnarray*}

The partial derivatives of the off-diagonal $\zeta$ function which
appear in the above expression can be computed from the spectral
representation in $D$ dimensions (${\bf x}\equiv(\tau,{\vec{z}})$,
$k=|{\bf k}|$)
\begin{eqnarray*}
\zeta_D(s;x,x')&=&\frac{2 (4\pi)^{-\frac{D-2}{2}}}
{\alpha\Gamma(\frac{D-2}{2})}
\int_0^\infty dk\, k^{D-3}\sum_{n=-\infty}^\infty\int_0^\infty
d\lambda\,\lambda [\lambda^2+k^2+m^2]^{-s}\nonumber\\
&&\times J_{\nu_n}(\lambda r')
J_{\nu_n}(\lambda r)e^{i{\bf k}\cdot({\bf x}-{\bf x}')+
i\frac{2\pi}{\alpha}n(\theta-\theta')},
\end{eqnarray*}
and then taking the coincidence limit. In this way one can easily show
that
\begin{eqnarray*}
\partial_\theta\partial_{\theta'}\zeta_D(s;x,x')|_{x=x'}&=&
-\partial_\theta^2\zeta_D(s;x,x')|_{x=x'}
\nonumber\\
\partial_{z_i}\partial_{{z_i}'}\zeta_D(s;x,x')|_{x=x'}&=&
-\partial_{z_i}^2\zeta_D(s;x,x')|_{x=x'}=
2\pi\zeta_{D+2}(s;x)\nonumber\\
\partial_\tau\partial_{\tau'}\zeta_D(s;x,x')|_{x=x'}&=&
-\partial_\tau^2\zeta_D(s;x,x')|_{x=x'}=2\pi\zeta_{D+2}(s;x)
\nonumber\\
\partial_{r'}\zeta_D(s;x,x')|_{x=x'}&=&\frac{1}{2}\partial_r 
\zeta_D(s;x).
\end{eqnarray*}
As far as $\partial_\theta^2\zeta_D(s;x,x')|_{x=x'}$ is concerned, 
in the $m=0$ case it is easy to see that we have
\begin{eqnarray*}
\partial_\theta^2\zeta_D^{m=0}(s;x,x')|_{x=x'}=
-\frac{r^{2s-D}\Gamma(s-\frac{D-1}{2})}
{(4\pi)^{\frac{D-2}{2}}\sqrt{\pi}\alpha\Gamma(s)}
H_\alpha(s-{\scriptstyle\frac{D-2}{2}}),
\end{eqnarray*}
where the function $H_\alpha(s)$ is defined and studied in the
Appendix {\bf A}. The massive case can then be treated using the
off-diagonal version of Eq. (\ref{general}) with the partial
coincidence limit $r=r'$, $z=z'$ and $t=t'$:
\begin{eqnarray*}
\partial_\theta^2\zeta_4(s;\theta,\theta')|_{\theta=\theta'}&=&
\frac{1}{2\pi i\Gamma(s)}\int \Gamma(z)
\partial_\theta^2\zeta_4^{m=0}(z;\theta,\theta')|_{\theta=\theta'}
\Gamma(s-z) m^{2z-2s}dz\nonumber\\
&=&\frac{-1}{2\pi i\Gamma(s)} \int_{\mbox{\scriptsize Re } z>3}
\frac{r^{2z-4}}{4\pi\sqrt{\pi}\alpha}
\Gamma(z-3/2)H_\alpha(z-1)\Gamma(s-z) m^{2z-2s}dz.\\
&=&-\frac{r^{2s-4}\mu^{2s}}{4\pi\sqrt\pi\alpha\Gamma(s)}\left\{
\sum_{n=0}^\infty\frac{(-1)^n}{n!}(mr)^{2n}\Gamma(s+n-3/2)
H_\alpha(s+n-1) \right.\nonumber\\
&&\left.\hspace{-2cm}+(mr)^{4-2s}
\sum_{n=1}^\infty\sum_{k=0}^\infty\frac{(-1)^k}{k!}
(mr)^{2\nu_n+2k}\nu_n^2\frac{\Gamma(s-\nu_n-k-2)
\Gamma(\nu_n+k+1/2)}
{\Gamma(2\nu_n+k+1)}\right\}.
\end{eqnarray*}
where, as usual, we have performed the integration shifting the
integration contour to the right and picking up the residues at
$z=s+n$ and $z=\nu_n+k+2$ to get an expansion in powers of $mr$
similar to Eq. (\ref{zetafun}).

As far as the second derivatives with respect to $r$ and $r'$ are
concerned, using the following identity, which can be proved using
some recursion formulas for the Bessel functions \cite{GR},
$$
[\partial_r J_\nu(\lambda r)]^2=-\frac{\nu^2}{r^2}J_\nu^2(\lambda r)
+\frac{1}{2r}\partial_r r\partial_r J_\nu^2(\lambda r)+\lambda^2
J_\nu^2(\lambda r),
$$
one can see that 
\begin{eqnarray*}
\partial_r\partial_{r'}\zeta_D(s;x,x')|_{x=x'}&=&
\frac{1}{2r}\partial_r r\partial_r\zeta_D(s;x)+\frac{1}{r^2}
\partial_\theta^2
\zeta_D(s;x,x')|_{x=x'}+\chi_D(s;x),\\
\partial^2_{r'}\zeta_D(s;x,x')|_{x=x'}&=&
-\frac{1}{2r}\partial_r\zeta_D(s;x)-\frac{1}{r^2}\partial_\theta^2
\zeta_D(s;x,x')|_{x=x'}-\chi_D(s;x),\\
\end{eqnarray*}
where the function $\chi_D(s;x)$ is defined in Appendix {\bf B}.
If $D=4$ and $m=0$ we simply have
\begin{eqnarray}
\chi_{D=4}^{m=0}(s;x)=4\pi(s-2)\zeta_{D=6}^{m=0}(s;x),
\label{chizero}
\end{eqnarray}
while the massive case is more complicate and is studied in the
Appendix.

Now we have all the pieces needed to compute the massive correction to
the energy-momentum tensor. For the actual calculation it is
convenient to follow \cite{BROT86} and \cite{ALO92}. We define the
renormalized stress tensor as
$$
\langle T_{\mu\nu}(x)\rangle_\alpha^R\equiv
\langle T_{\mu\nu}(x)\rangle_\alpha-
\langle T_{\mu\nu}(x)\rangle_{2\pi}
$$
which, considering only the first massive correction, turns out to be
\begin{eqnarray*}
\langle T_{\theta\theta}(x)\rangle_\alpha^R&=&
\frac{1}{4\pi\alpha r^2}\left[2H_\alpha(0)-(6\xi-1)I_\alpha(0)+
(mr)^2\left(H_\alpha(1)+\xi\frac{\nu-1}{\nu}\right)\right],\\
\langle T_{rr}(x)\rangle_\alpha^R&=&
\frac{-1}{4\pi\alpha r^4}
\left[2H_\alpha(0)-I_\alpha(-1)-(2\xi-1)I_\alpha(0)+
(mr)^2\left(H_\alpha(1)+\xi\frac{\nu-1}{\nu}\right)\right],\\ 
\langle T_{tt}(x)\rangle_\alpha^R&=& \frac{-1}{8\pi\alpha
r^4}\left[I_\alpha(-1)+2(4\xi-1)I_\alpha(0)+
(mr)^2I_\alpha(0)\right],\\ 
\langle T_{zz}(x)\rangle_\alpha^R
&=&\langle T_{tt}(x)\rangle_\alpha^R,
\end{eqnarray*}
The tensor can be written in the more familiar form
\begin{eqnarray}
\langle {T_\mu}^\nu(x)\rangle_\alpha^R&=&
\frac{-1}{1440\pi^2 r^4}\left [\left(\nu^4-1\right)
\mbox{diag}(1,1,-3,1)+10(6\xi-1)\left(\nu^2-1\right)
\mbox{diag}(2,-1,3,2)\right.\nonumber\\
&&+15(mr)^2(\nu-1)
(12\xi-1-\nu) \mbox{diag}(0,1,-1,0)\nonumber\\
&&\left.-15(mr)^2\left(\nu^2-1\right)
\mbox{diag}(1,0,0,1)\right].
\label{stress}
\end{eqnarray}
In the limit $m\rightarrow 0$ the result is in agreement with that 
obtained by others authors \cite{DO87a,FRSE87}.
The components $\langle T_{rr}(x)\rangle_\alpha^R$ and
$\langle T_{\theta\theta}(x)\rangle_\alpha^R$ satisfy the
equation
$$
\frac{d}{dr}(r {T^r}_r)={T^\theta}_\theta,
$$
which follows from the conservation law $\nabla_\mu {T^\mu}_\nu=0$. 
It is interesting to note that in the corrections to the stress tensor
of order $(mr)^2$  are not present the logarithmic divergences which
are instead present in the vacuum fluctuations at the same order (see
Eq. (\ref{vacuum})). The logarithmic terms appear in higher order
corrections, which can also be computed with the method we have
developed. Actually only the logarithms in the term of order $(mr)^4$
give rise to divergences, since at higher orders they are multiplied
by positive powers of $r$. Indeed, the correction of order $(mr)^4$ to
$\langle T_{\theta\theta}(x)\rangle_\alpha^R$ is
\begin{eqnarray*}
\frac{1}{4\pi\alpha r^2}\left\{-\frac{(mr)^4}{8}
\left[2\frac{\ln\nu+1}{\nu}+(2\xi-1)A(r)+4\xi-1\right]\right.\\
\left.+(mr)^{2+2\nu}\nu[2\xi+(4\xi-1)\nu]\frac{\Gamma(-\nu-1)
\Gamma(\nu+1/2)}{\sqrt{\pi}\Gamma(2\nu+1)}\right\},
\end{eqnarray*}
from which one can also compute the
correction to $\langle T_{rr}(x)\rangle_\alpha^R$
by means of the conservation law. The correction
to $\langle T_{tt}(x)\rangle_\alpha^R=
\langle T_{zz}(x)\rangle_\alpha^R$ is 
\begin{eqnarray*}
-\frac{1}{8\pi\alpha r^4}\left\{\frac{(mr)^4}{4}
\left[2B(r)-(4\xi-1)(A(r)+1)-\frac{1}{\nu}\right]\right. \\
\left.+(mr)^{2+2\nu}2(4\xi-1)\nu^2\frac{\Gamma(-\nu-1)
\Gamma(\nu+1/2)}
{\sqrt{\pi}\Gamma(2\nu+1)}\right\}.
\end{eqnarray*}
In the above equations we have set
\begin{eqnarray*}
A(r)&=&2G_\alpha(2)+2\gamma+2\ln\frac{mr}{2}.\\
B(r)&=&\frac{\nu-1}{\nu}\left(\ln\frac{mr}{2}+
\gamma-\frac{1}{2}\right)+\frac{\ln\nu}{\nu}.
\end{eqnarray*}

An interesting point to be discussed is the dependence on the
parameter $\xi$ which fixes the coupling of the scalar field with the
gravity. In the introduction we said that we would consider the
minimally coupled case only, $\xi=0$. This because it has been shown
in \cite{AlKaOt96} that the idealized conical space is a good model of
the space time of a cosmic string  only for the minimally coupled
case. For  nonminimally coupled  fields quantities like $\langle
\phi^2(x)\rangle$ will depend on the details of the metric in the core
of the string even very far from the string. However, one could be
interested, e.g., in  finite-temperature fields in the Rindler space,
where there is a true conical singularity and  not just an
idealization of a non-singular metric. In these cases it is
interesting to consider also nonminimally coupled fields.

{}From the mathematical point of view, it is not clear which is the
meaning of the field equation $[-\Delta+m^2+\xi R]\phi=0$ when the
curvature $R$ has Dirac's delta singularities, and our choice of
$\xi=0$ allowed us to avoid the problem. A possible way to  define the
problem is to smooth out the singularity, as done in
\cite{AlOt90,ALO92,FUSOL95}: as a result, one finds that when the
regularization of the singularity is removed to recover the conical
space also the dependence of the Green's function on the parameter
$\xi$ vanishes \cite{ALO92}. Then we can argue that also the $\zeta$
function is independent of $\xi$, since the modes used to construct
the $\zeta$ function are essentially the same as those for the Green's
function.

The dependence on the parameter $\xi$ comes back into play when
considering the stress tensor. It is worth while noticing that even
when the manifold if flat everywhere (in our case there is a Dirac's
delta singularity in the curvature at the core of the string) the
parameter $\xi$ remains in the theory as a relic of the fact that
$T_{\mu\nu}$ is obtained by varying the metric $g_{\mu\nu}$ in the
field Lagrangian \cite{birrel}. One can then see that, if $R=0$,  the
global conserved quantities as total energy should not depend on the
value of $\xi$. This is  because the contributions to those quantities
due to $\xi$ are discarded into boundary surface integrals which
generally vanish. However, this is not the case dealing with the
cosmic string because such integrals diverge at the origin. Notice
that similar problems appear working in subregions of the Minkowski
space in presence of boundary conditions \cite{birrel}. In the
point-splitting approach, the stress tensor is obtained applying the
$\xi$-dependent operator $D_{\mu\nu}(x,x')$ (see Eq. (\ref{dimunu}))
to the Green's function or to the $\zeta$ function. Therefore, if the
above argument that the Green's function  (or the $\zeta$ function) is
independent of $\xi$ holds, it is not contradictory to compute the
energy-momentum tensor applying $D_{\mu\nu}(x,x')$ with $\xi\neq 0$ to
the the Green's function computed setting $\xi=0$, as we have done
above and as done by most authors.

\begin{center}
{\large\bf V. Conclusions}
\end{center}

In this paper we have studied the $\zeta$ function of a massive scalar
field in a cosmic string background, and we have obtained an
expression, Eq. (\ref{zetafun}), which is useful in the region near the
core of the string, $mr\ll 1$. By means of this  expression 
we have  computed the massive corrections to the vacuum
fluctuations, Eq. (\ref{vacuum}), and to the energy-momentum tensor,
Eq. (\ref{stress}), up to order $(mr)^4$, going beyond the known
results. Higher corrections are also computable.

Possible extensions of this paper are the inclusion of a magnetic
flux carried by the string, which gives Aharonov-Bohm effects, and 
the case of spin-$1/2$ fields.  Also the limit $mr\gg 1$ is worth
studying, for example rewriting the $\zeta$ function (\ref{zetafun})
in terms of the hypergeometric function ${}_1F_2[a;b,c;z]$ and 
then using its asymptotic behaviour for large $z$. However, 
the vacuum fluctuations and the energy-momentum tensor 
in this limit have been already obtained by others authors
with different methods \cite{ShiraHire92,GuimLin94}.

As a possible application of the results of this paper we point out the
study of the backreaction of the energy momentum tensor of a 
massive scalar field on the background metric of the cosmic string, 
as done by Hiscock \cite{Hiscock87} and Guimar{\~a}es \cite{MEX97}.
The computation seems feasible since the first massive correction to
the energy momentum tensor does not involve logarithmic terms. 
We hope to cover this topic in a future paper.
\\

\noindent {\large\bf Acknowledgments}

\noindent I would like to thank Valter Moretti and, in particular,
Sergio Zerbini for useful discussions and suggestions.

\vspace{1cm}

\begin{center}
{\large\bf Appendix A}
\end{center}

In this Appendix we study the function $H_\alpha(s)$, which is 
defined as the analytic continuation of following series:
\begin{eqnarray*}
H_\alpha(s)=\sum_{n=1}^\infty\nu_n^2
\frac{\Gamma(\nu_n-s+1)}{\Gamma(\nu_n+s)},
\hspace{1cm}\nu_n=\frac{2\pi}{\alpha}|n|.
\end{eqnarray*}
It can be studied and analytically continued proceeding exactly
as for the function $G_\alpha(s)$ studied in the Appendix of \cite{ZCV}.
Then one sees that the series converges for $\mbox{Re} s>2$ and that
the analytic continuation for $[n/2]<\mbox{Re} s<2$ (here $[n/2]$
represents the integer part of $n/2$) is given by
\begin{eqnarray*}
\sum_{j=0}^{[n/2]+1}c_j(s)\nu^{3-2s-2j}\zeta_R(2s+2j-3)
+\sum_{k=1}^\infty \nu_n^2 f_n(\nu_k,s),
\end{eqnarray*}
where $\zeta_R$ is the Riemann $\zeta$ function and the function
$f_n(\nu,s)$ is generally unknown, but vanishes for $s=1/2,0,-1/2,
-1,\dots$. The coefficients $c_j(s)$ vanish for $s=-n/2$,
($n=-1,0,1,2,\dots$) for all $j>(n+1)/2$, and the first ones 
have been given in \cite{ZCV}.

The function $H_\alpha(s)$ has then a simple pole in $s=2$ and 
near this pole we have 
\begin{eqnarray*} 
H_\alpha(s)=\frac{1}{2\nu(s-2)}+\frac{1}{\nu}(\gamma-\ln\nu)+
{\cal O}\left((\nu-2)^2\right).
\end{eqnarray*}
Moreover, it has simple poles at $s=1+\nu_n+k$, $k=1,2,\dots$, due to
the gamma function in the numerator of the terms of the series, with
obvious residue. Finally, it is possible to compute the value of the
function $H_\alpha(s)$ for some useful value of $s$:
\begin{eqnarray*}
H_\alpha(0)&=&\frac{1}{120\nu}(\nu^4-1),\nonumber\\
H_\alpha(1)&=&-\frac{1}{12\nu}(\nu^2-1).
\end{eqnarray*}

\begin{center}
{\large\bf Appendix B}
\end{center}

In this Appendix we study the function $\chi_D(s;x)$, which is defined
by the analytic continuation of the following spectral representation
\begin{eqnarray}
\chi_D(s;x)=\frac{2 (4\pi)^{-\frac{D-2}{2}}}{\alpha\Gamma((D-2) /2)}
\int_0^\infty dk\, k^{D-3}\sum_{n=-\infty}^\infty\int_0^\infty
d\lambda\,\lambda^3 [\lambda^2+k^2+m^2]^{-s}J_{\nu_n}^2(\lambda r),
\label{chi}
\end{eqnarray}
which is the same as $\zeta_D(s;x)$ with $d\lambda\,\lambda
\rightarrow d\lambda\,\lambda^3$. Note that the massless case 
is trivial, since in that case the function $\chi$ is simply related
to the $\zeta$ function, see Eq. (\ref{chizero}). In the massive case,
we can proceed in analogy to the massive $\zeta$ function, namely
employing Eq. (\ref{general}): considering $D=4$ and using Eq.
(\ref{chizero})
\begin{eqnarray*}
\chi^{m}_{D=4}(s;x)&=&\frac{1}{2\pi i \Gamma(s)}
\int_{\sigma-i\infty}^{\sigma-i\infty}
\Gamma(z)\chi^{m=0}_{D=4}(z;x)m^{2z-2s}\Gamma(s-z) dz
\nonumber\\
&=&\frac{4\pi}{2\pi i \Gamma(s)}
\int_{\sigma-i\infty}^{\sigma-i\infty}
\Gamma(z)(z-s)\zeta^{m=0}_{D=6}(z;x)m^{2z-2s}\Gamma(s-z) dz.
\end{eqnarray*}
Then we go on as usual splitting $\zeta^{m=0}_{D=6}(z;x)$
as $\zeta=\zeta_<+\zeta_>$ and performing the integrals 
shifting the integration contour to the left and picking up
the residues of the poles. The final result is an expansion
in powers of $mr$:
\begin{eqnarray}
\chi^{m}_{D=4}(s;x)=\frac{r^{2s-6}}{4\pi\alpha\Gamma(s)}
\left\{\sum_{n=0}^\infty\frac{(-1)^n}{n!}(s+n-2)(mr)^{2n}
I_\alpha(s+n-2)
\right.\nonumber\\
+\sum_{n=0}^\infty\frac{(-1)^n}{n!}(n+1)(mr)^{6+2n-2s}
\frac{\Gamma(n+1/2)\Gamma(s-n-3)}{2\sqrt{\pi}\Gamma(n+1)}
\nonumber\\
\left.+(mr)^{6-2s}
\sum_{n=1}^\infty\sum_{k=0}^\infty
\frac{(-1)^k}{k!}(\nu_n+k+1)(mr)^{2\nu_n+2k}
\frac{\Gamma(\nu_n+k+1/2)\Gamma(s-\nu_k-k-3)}{2\sqrt{\pi}
\Gamma(2\nu_n+k+1)}\right\}.
\nonumber\\
&&
\label{chifinal}
\end{eqnarray}

\end{document}